\documentclass[oldversion]{aa}
\usepackage{graphics, epsfig}
\begin{document}

\title{Structure formation and CMBR anisotropy spectrum in the inflessence model}

\author{A.A. Sen\inst{1}, V.F. Cardone\inst{2}, S. Capozziello\inst{3}, \and A. Troisi\inst{3}}

\offprints{V.F. Cardone, \email{winny@na.infn.it}}

\institute{Department of Physics and Astronomy, Vanderbilt
University, Nashville, TN 37235, USA \and Dipartimento di Fisica
``E.R. Caianiello'', Universit{\`{a}} di Salerno and INFN, Sezione
di Napoli, Gruppo Collegato di Salerno, Via S. Allende, 84081 -
Baronissi (Salerno), Italy \and Dipartimento di Scienze Fisiche,
Universit\`a di Napoli and INFN, Sezione di Napoli, Compl. Univ.
di Monte S. Angelo, Edificio G, Via Cinthia, 80121 - Napoli,
Italy}

\date{Received / Accepted}

\abstract{The inflessence model has recently been proposed in an
attempt to explain both early inflation and present day
accelerated expansion within a single mechanism. The model has
been successfully tested against the Hubble diagram of Type Ia
Supernovae, the shift parameter, and the acoustic peak parameter.
As a further mandatory test, we investigate here structure
formation in the inflessence model determining the evolution of
matter density contrast $\delta \equiv \delta \rho_M/\rho_M$ in
the linear regime. We compare the growth factor $D(a) \equiv
\delta/a$ and the growth index $f(z) \equiv d\ln{\delta}/d\ln{a}$
to these same quantities for the successful concordance
$\Lambda$CDM model with a particular emphasis on the role of the
inflessence parameters $(\gamma, z_Q)$. We also evaluate the
anisotropy spectrum of the cosmic microwave background radiation
(CMBR) to check whether the inflessence model may be in agreement
with the observations. We find that, for large values of $(\gamma,
z_Q)$, structure formation proceeds in a similar way to that in
the $\Lambda$CDM scenario, and it is also possible to nicely fit
the CMBR spectrum.

}

\titlerunning{Growth of structures in the inflessence model}

\maketitle

\section{Introduction}

It is now widely accepted that we live in a spatially flat
universe undergoing an accelerated expansion and made out of $\sim
95\%$ dark ingredients which we know little about. On the one
hand, observations of the CMBR anisotropy spectrum (see, e.g.,
\cite{Boom}; \cite{Max}; \cite{WMAP}; and \cite{page} for a
review) indicates that the total energy density attains the
critical one so that the universe is spatially flat. On the other
hand, the SNeIa Hubble diagram (\cite{Gold}; \cite{SNLS}) is a
clear signature of the cosmic speed\,-\,up of the universe
expansion, hence discarding with a great degree of confidence the
old standard picture of a matter\,-\,dominated universe. Finally,
the matter power spectrum and the clustering properties of
galaxies observed in large galaxy surveys (\cite{Pope04};
\cite{Cole05}) point towards the existence of dark matter
suggesting that its density parameter $\Omega_M$ is of the order
of 0.3, far lower than the SCDM value $\Omega_M = 1$, thus
stressing the need of a further component to achieve the critical
density. When combined together, this impressive set of
observations motivates the entrance on the scene of a new player
dominating the energy budget and driving the accelerated
expansion. This elusive and mysterious component is referred to as
dark energy.

Although the need for dark energy is clear, its nature and
fundamental properties are completely unknown. The simplest
candidate is the well\,-\,known cosmological constant $\Lambda$
(\cite{Carrol}; \cite{Sahni}), which perfectly matches a wide
range of observations (\cite{Teg03}; \cite{Sel04}), hence awarding
the name of concordance model to the scenario based on $\Lambda$
and cold dark matter (CDM). Despite this impressive success, the
$\Lambda$CDM model is plagued by serious theoretical shortcomings,
thus motivating the search for alternative schemes. This has
opened the way to an overwhelming flood of papers proposing
different models for explaining the cosmic speed\,-\,up and the
CMBR anisotropy spectrum with proposals ranging from a dynamical
$\Lambda$ originating from a scalar field (dubbed quintessence)
rolling down its self\,-\,interaction potential (see, e.g.,
\cite{PR03} and \cite{Pad03} for comprehensive reviews), to
unified models of dark matter and dark energy such as the
Chaplygin gas (\cite{KMP01}; \cite{BTV02}; \cite{BBS03}) and the
Hobbit models (\cite{Hobbit}), to braneworld inspired scenarios
(\cite{DGP}; \cite{LSS04}) and higher order theories of gravity
both in the metric (\cite{capozcurv}; \cite{noiijmpd};
\cite{ND03}; \cite{CDTT}; \cite{mimic}) and the Palatini
(\cite{V03}; \cite{MW03}; \cite{Flanagan}; \cite{Allemandi};
\cite{noifranc}, \cite{multa}) formulations. Although radically
different in their theoretical aspects, all of these models are
equally viable from the observational point of view, thus
indicating that better quality data, higher redshift probes, or
new tests are in order to break some of the degeneracies among
different models.

It is worth noting that both current theoretical schemes and
observational evidences predict that the evolutionary history of
the universe comprises two periods of accelerated expansion,
namely the inflationary epoch and the present day dark energy
dominated phase. In both cases, the expansion is usually
interpreted as the result of the presence of a negative pressure
fluid dominating the energy budget. It is natural to wonder
whether a single (effective) fluid may indeed be responsable for
both periods of accelerated expansion. At the same time, this
fluid should be subdominant during the radiation and matter
dominated epochs so as not to interfere with baryogenesis and
structure formation. While it is quite difficult to theoretically
formulate the properties of such a fluid, it is, on the contrary,
clear what its phenomenological features are. Inspired by these
considerations, some of us have recently proposed the {\it
inflessence} model (\cite{ie}). Based on a suitable Ansatz for the
dependence of the energy density on the scale factor $a$, the
inflessence scenario has been shown to be able to fit the SNeIa
Hubble diagram, and also give correct values for the shift
${\cal{R}}$ (\cite{Bond,defR}) and acoustic peak ${\cal{A}}$
(\cite{Eis05}) parameters. While this result gives an
observational motivation for the model, inflessence is also well
founded theoretically, since it can be interpreted both in terms
of scalar field quintessence and as an effective model coming from
fourth order theories of gravity.

Motivated by these observational and theoretical results, we
extend the analysis of the inflessence model here by investigating
structure formation in this scenario. Moreover, we also present a
preliminary analysis of the CMBR anisotropy spectrum. Both these
features are standard observables in cosmology nowadays, and it is
therefore mandatory to check whether the inflessence model is able
to survive these tests.

The structure of the paper is as follows. Section\,2 briefly
recalls the main features of the inflessence model and explains
what the roles played by its characterizing parameters are. In
Sect.\,3, the evolution of matter density perturbations is studied
in the linear regime, assuming that the inflessence fluid does not
cluster on the subhorizon scale, which is indeed the case of most
dark energy models. Section\,4 is dedicated to a discussion of how
the growth index depends on the inflessence parameters and the
constraints that could possibly be extracted from a precise
determination of this quantity. The CMBR anisotropy spectrum is
evaluated in Sect.\,5, while a summary of the results and the
conclusions are presented in Sect.\,6.

\section{The inflessence model}

The key ingredient is the following Ansatz for the inflessence
energy density\,:

\begin{equation}
\rho(a) = {\cal{N}} a^{-3} \left ( 1 + \frac{a_I}{a} \right
)^{\beta} \left ( 1 + \frac{a}{a_Q} \right )^{\gamma} \label{eq:
rhovsa}
\end{equation}
with a normalization constant ${\cal{N}}$, slope parameters
$(\beta, \gamma)$, and two scaling values of the scale factor $a_I
<< a_Q$. For later applications, it is convenient to rewrite
Eq.(\ref{eq: rhovsa}) in terms of the redshift $z = 1/a - 1$
(having set $a_0 = 1$ with the subscript $0$ denoting henceforth
quantities evaluated at the present day, i.e. $z = 0$)\,:

\begin{equation}
\rho(z) = {\cal{N}} (1 + z)^3 \left ( 1 + \frac{1 + z}{1 + z_I}
\right )^{\beta} \left ( 1 + \frac{1 + z_Q}{1 + z}\right
)^{\gamma} \ , \label{eq: rhovsz}
\end{equation}
having defined\,:

\begin{equation}
z_I = 1/a_I - 1 \ , \label{eq: defzi}
\end{equation}
\begin{equation}
z_Q = 1/a_Q - 1 \ . \label{eq: defzq}
\end{equation}
From Eq.(\ref{eq: rhovsa}), it is quite easy to see that the
energy density of the inflessence fluid scales like that of dust
matter $(\rho \sim a^{-3}$) in the range $a_I << a << a_Q$, so
that, given typical values for $(a_I, a_Q)$, the fluid follows
matter for a large part of the universe history, while it scales
differently only during the very beginning ($a << a_I$) and the
present ($a >> a_Q$) periods. Moreover, choosing $\beta = -3$, the
fluid energy density remains constant for $a << a_I$, thus
behaving like the usual cosmological constant $\Lambda$ during the
early epoch of the universe evolution. Finally, the slope
parameter $\gamma$ determines how the fluid energy density scales
with $a$ in the present epoch.

It is still more instructive to look at the equation of state
(EoS) $w \equiv p/\rho$, where $p$ is the fluid pressure. Using
the continuity equation\,:

\begin{equation}
\dot{\rho} + 3 H (\rho + p) = 0 \ , \label{eq: cont}
\end{equation}
with the Hubble parameter $H = \dot{a}/a$ and inserting
Eq.(\ref{eq: rhovsa}) into Eq.(\ref{eq: cont}), after some algebra
we get\,:

\begin{equation}
w = \frac{\beta}{3} \left ( \frac{1 + z}{2 + z + z_I} \right ) -
\frac{\gamma}{3} \left ( \frac{1 + z_Q}{2 + z + z_Q} \right ) \ .
\label{eq: wvsz}
\end{equation}
It is worth noting that $w$ does not depend either on $\gamma$ or
on $z_Q$ for high values of $z$, which is to be expected looking
at Eq.(\ref{eq: wvsz}). On the contrary, these two parameters play
a key role in determining the behavior of the EoS over the
redshift range $(0, 100)$, which represents most of the history of
the universe (in terms of time).

The role of the different quantities $(\beta, \gamma, z_I, z_Q)$
is better understood considering the asymptotic limits of the EOS.
We easily get\,:

\begin{equation}
\lim_{z \rightarrow \infty}{w(z)} = \frac{\beta}{3} \ , \label{eq:
wlim}
\end{equation}
which shows that setting $\beta = -3$, the fluid EoS approaches
that of the cosmological constant, i.e., $w_{\Lambda} = -1$, in
the very early universe. In general, if we impose the constraint
$\beta < -1$, we get a fluid having a negative pressure in the far
past so that it is able to drive the accelerated expansion
occurring during the inflationary epoch.  It is therefore clear
that $z_I$ controls the transition towards the past asymptotic
value, in the sense that the larger is $z$ with respect to $z_I$,
the smaller is the difference between $w(z)$ and its asymptotic
limit $\beta/3$. This consideration suggests that $z_I$ has to
take quite high values (indeed, far greater than $10^3$) since,
for $z >> z_I$, the universe is in its inflationary phase.

In the asymptotic future (i.e., $z \rightarrow -1$), we get\,:

\begin{equation}
\lim_{z \rightarrow -1}{w(z)} = - \frac{\gamma}{3} \label{eq:
wlimbis}
\end{equation}
so that the slope parameter $\gamma$ determines the future
evolution of the universe. For instance, for $\gamma = 3$, the
universe finally ends in a de Sitter state (as for the concordance
$\Lambda$CDM model), while a Big Rip occurs for $\gamma > 3$, as
in phantom models (\cite{phantom1}; \cite{phantom2}).

Let us now consider the present day value of $w$ that turns out to
be\,:
\begin{equation}
w_0 = \frac{\beta}{3 (2 + z_I)} - \frac{\gamma}{3} \left ( \frac{1
+ z_Q}{2 + z_Q} \right ) \simeq -\frac{\gamma}{3} \left ( \frac{1
+ z_Q}{2 + z_Q} \right ) \label{eq: wz}
\end{equation}
where, in the second line, we used the fact that $z_I$ is very
large. Given that $z_Q > 0$, to have a present day accelerated
expansion, $w_0$ should be lower than $-1/3$ so that we get the
constraint $\gamma > (2 + z_Q)/(1 + z_Q)$. Moreover, depending on
the values of $\gamma$ and $z_Q$, $w_0$ could also be smaller than
$w_{\Lambda}$ so that we may recover phantom\,-\,like models. The
parameter $z_Q$ then regulates the transition to the dark
energy\,-\,like dominated period.

In summary, the inflessence fluid with energy density and EOS
given by Eqs.(\ref{eq: rhovsz}) and (\ref{eq: wvsz}) is able to
drive the accelerated expansion of the universe during both the
inflationary epoch and the present day period. Therefore, such a
fluid plays the role of both the {\it inflaton} and the {\it
quintessence} scalar field, hence the motivation for the name {\it
inflessence} (contracting the words {\it inflationary
quintessence}).

A general comment is in order here. Although originally proposed
as a phenomenological Ansatz for the energy density, the
inflessence model could also be seen as an effective
parametrization of the scale factor during the universe's
expansion, which translates into the above scaling of $\rho(z)$
given the known behavior of the matter and radiation components.
This equivalent representation is particularly interesting during
the inflationary epoch. Indeed, if the inflessence fluid plays the
role of the inflaton field, one should wonder whether its
self\,-\,interaction potential is able to give rise to reheating.
Discussing this issue is outside our aims, but we stress that
considering the model as a parametrization of the scale factor
rather than the energy density makes it possible to escape
problems with reheating.

It is worth noting that, since $\rho$ scales with $a$ as the dust
matter energy density for a long period of the universe history,
the coincidence problem is partially alleviated. Indeed, the dark
energy and the matter components track each other for a long
period so that their near equality today turns out to be a
consequence of the relatively recent change of the scaling of the
inflessence energy density with $a$. However, there is still a
certain degree of fine tuning since the model parameters have to
be set in a suitable way so that the transition from decelerated
to accelerated expansion takes place only recently. Moreover,
although $w_0 < -1$ is possible depending on the values of
$(\gamma, z_Q)$, the possibility to avoid the Big Rip still
remains if $\gamma \le 3$, although such low values seem to be
disfavored by the fitting to the dimensionless coordinate distance
of SNeIa and radiogalaxies.

As a final important remark, let us stress that, although
phenomenologically inspired, the inflessence model may also be
theoretically well\,-\,founded. Indeed, as shown in Cardone et al.
(2005), one could obtain Eq.(\ref{eq: rhovsz}) as a result of
scalar field quintessence with a self\,-\,interacting potential,
which can be well approximated as a simple power law with negative
slope. As an alternative explanation, the inflessence model may
also be recovered as the effective fluid like description of a
fourth order $f(R)$ theory in which the Einsteinian gravity
Lagrangian $f(R) \propto R$ is replaced by a suitably
reconstructed $f(R)$. From a different point of view, this also
offers the possibility of considering the inflessence scenario as
an analytical parametrization for studying a wide class of diverse
models. As such, investigating the growth of structures for the
inflessence fluid gives indirect constraints on these models too.
To this end, it is worth stressing that we are assuming that the
inflessence model may be recovered as a particular case of scalar
field quintessence so that the gravity Lagrangian is the standard
one, and we can therefore resort to the usual perturbation theory
for our analysis of the growth of structures. Should we choose the
interpretation of the inflessence model in the framework of $f(R)$
theories, we should adopt a different formalism.

\section{Linear growth of fluctuations}

The inflessence model has been successfully tested against the
SNeIa Hubble diagram, also taking into account the shift parameter
(which essentially gives the distance to the last scattering
surface) and the baryonic acoustic peak parameter. Although
successfully fitting this dataset is mandatory for any realistic
dark energy model, such a test only probes the dynamics of the
background cosmology. Indeed, dark energy also has impacts on the
evolution of density perturbations.

As is well known, the universe is homogenous and isotropic only on
the largest scales. As a consequence, while one can still use the
standard FRW description when considering the dynamics of the
universe on the scales of interest, smaller scale evolution must
take into account the inhomogeneities of the spacetime. Using the
equations of motion for this perturbed metric, one can derive the
growth of density perturbations. Moreover, it is possible to
demonstrate that, because of its high sound speed, dark energy
only clusters on scales that are far larger than those of galaxies
and galaxy clusters. As a consequence, dark energy affects the
structure formation process only because of its background energy
density, which concurs to determine the expansion rate. This is
indeed also the case for the inflessence model we are considering
here, so that we may resort to the standard theory to investigate
the growth rate of matter perturbations in the linear regime.

Denoting the matter density contrast with $\delta \equiv \delta
\rho_M/\rho_M$, the perturbation equation reads\,:

\begin{equation}
\ddot{\delta} + 2 H \dot{\delta} - 4 \pi G \rho_M \delta = 0 \ .
\label{eq: perttime}
\end{equation}
It is convenient to change the variable from $t$ to the scale
factor $a$ so that Eq. (\ref{eq: perttime}) may finally be
rewritten as:

\begin{equation}
\delta^{\prime \prime} + \left [  \frac{3}{a} +
\frac{(\ln{E^2})^{\prime}}{2} \right ] \delta^{\prime} - \frac{3
\Omega_M}{2 E^2 a^{5}} \delta = 0 \ , \label{eq: pertscale}
\end{equation}
where the prime denotes the derivation with respect to the scale
factor $a$ and we have used $4 \pi G \rho_M = (3/2) \Omega_M H_0^2
a^{-3}$ and defined $E^2 = H^2/H_0^2$. To study the evolution of
perturbations in the inflessence scenario, we have only to insert
into Eq.(\ref{eq: pertscale}) the corresponding expression for
$E^2$, which reads\,:

\begin{equation}
E^2(a) = H^2(a)/H_0^2 = \Omega_r a^{-4} + \Omega_M a^{-3} +
\Omega_X g(a) \ , \label{eq: ea}
\end{equation}
where $\Omega_r$, $\Omega_M$, and $\Omega_X$ are the present day
values of the density parameters for radiation, dust matter, and
inflessence, respectively, and\,:

\begin{equation}
g(a) = a^{-3} \left ( \frac{1 + a_I/a}{1 + a_I} \right )^{\beta}
\left ( \frac{1 + a/a_Q}{1 + 1/a_Q} \right )^{\gamma} \ .
\label{eq: ga}
\end{equation}
Note that, consistent with the position of the first peak in the
CMBR anisotropy spectrum, we have assumed a spatially flat
universe so that $\Omega_r + \Omega_M + \Omega_X = 1$, although
some slight deviations from spatial flatness are still allowed by
the data when a time\,-\,varying dark energy equation of state is
used.

Since for a matter\,-\,only universe $\delta \propto a$, it is
useful for studying the effect of dark energy to divide this
behavior out and switch to the growth variable $D \equiv
\delta/a$. Starting from Eq.(\ref{eq: pertscale}), it is quite
easy to determine the equation governing the evolution of this
latter quantity\footnote{Actually, one may also use $\ln{a}$
instead of $a$ as expansion variable. See Linder (2005) for
different equivalent equations to determine $D(a)$.}\,:

\begin{eqnarray}
D^{\prime \prime} & +  & \left [ \frac{5}{a} +
\frac{(\ln{E^2})^{\prime}}{2} \right ] D^{\prime} \nonumber \\
~ & + &  \left [ \frac{3}{a} \left ( 1 - \frac{\Omega_M}{2 E^2
a^3} \right ) + \frac{(\ln{E^2})^{\prime}}{2} \right ] \frac{D}{a}
= 0 \ . \label{eq: eqpertd}
\end{eqnarray}
Equation (\ref{eq: eqpertd}) may be solved analytically only in
very special cases (see, e.g., \cite{perc} and references
therein), while, for our model (and indeed for most of dark energy
models), we have to resort to numerical integration using the
boundary conditions $\dot{D} = 0$ and $D = 1$ as $a \rightarrow
0$. Actually, it is not necessary to integrate from $a = 0$, but
one may set $a = a_{LS} = (1 + z_{LS})^{-1}$ as the initial
condition, $z_{LS}$ being the redshift of the last scattering
surface that we compute using the approximated relation in Hu \&
Sugyiama (1996). To this end, we set $\omega_b = \Omega_b h^2 =
0.0214$ in accordance with the nucleosynthesis constraints
(\cite{Kirk}) and $h = 0.664$ consistent with the Hubble diagram
of low redshift SNeIa (\cite{DD04}).

\begin{figure}
\centering \resizebox{8.5cm}{!}{\includegraphics{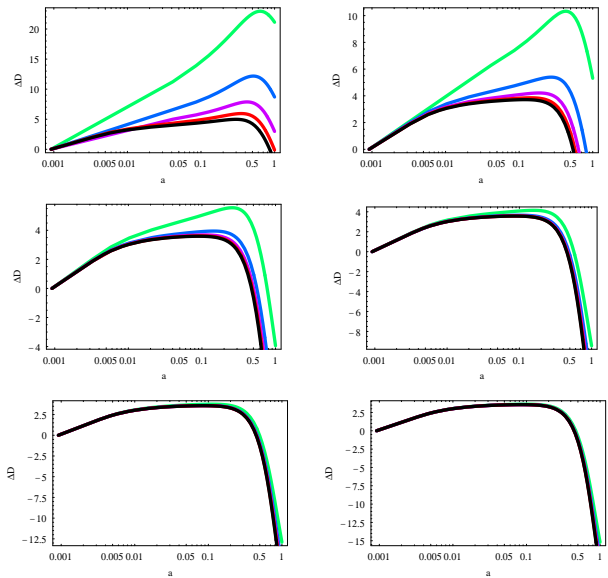}}
\caption{$\Delta D$ vs. $a$ for models with $\gamma$ from $3.5$ to
$8.5$ (in steps of $1$) from the top left to the bottom right
panel, and $z_Q$ from $1.0$ to $5.0$ (in steps of $1$) from the
uppermost to the lowermost curve. The other parameters are set as
explained in the text.} \label{fig: figone}
\end{figure}

Rather than looking at $D(a)$ directly, it is more interesting to
consider the quantity $\Delta D \equiv 1 - D/D_{\Lambda}$, which
represents the percentage deviation of the growth factor for the
inflessence model with respect to that for the concordance
$\Lambda$CDM one. Figure \ref{fig: figone} shows $\Delta D$
(multiplied by 100 for sake of clarity) as a function of the scale
factor $a$ for different combinations of the inflessence
parameters $(\gamma, z_Q)$, having set ($\Omega_M, \Omega_r) =
(0.28, 9.89 {\times} 10^{-5})$ (for both the inflessence and the
$\Lambda$CDM model) and fixed $(\beta, z_I)$ to their fiducial
values $(-3, 3454)$ (\cite{ie}). Note that these latter parameters
play a negligible role in our analysis since they mainly affect
the evolution of the fluid in the very early inflationary epoch.
It is worth stressing that setting $z_I = 3454$ does not at all
mean that we are assuming that inflation took place for $z$ near
this value. On the contrary, as could be easily checked, the
universe undergoes inflation only for $z >> z_I$ so that the exact
value of this latter parameter does not set the end of any
inflationary period, which could lead to possible problems with
nucleosynthesis.

Not surprisingly, the evolution of the growth factor highly
depends on the values of the parameters $(\gamma, z_Q)$, and both
negative and positive deviations from the growth factor in the
$\Lambda$CDM model may be obtained. Nevertheless, some general
results may be inferred. First, we note that, although deviations
as large as $20\%$ may be obtained, for most of the parameter
space $(\gamma, z_Q)$ the growth factor of the inflessence model
is comfortably similar to the $\Lambda$CDM one over the range $0.5
\le a \le 1$, i.e., $0 \le z \le 2$, where structure formation
mainly occurs. Although detailed numerical simulations should be
performed, this preliminary result makes us confident that the
assembly of galaxies and clusters of galaxies should have taken
place in a way that is quite similar to the one in the
$\Lambda$CDM model.

\begin{figure}
\centering \resizebox{8.5cm}{!}{\includegraphics{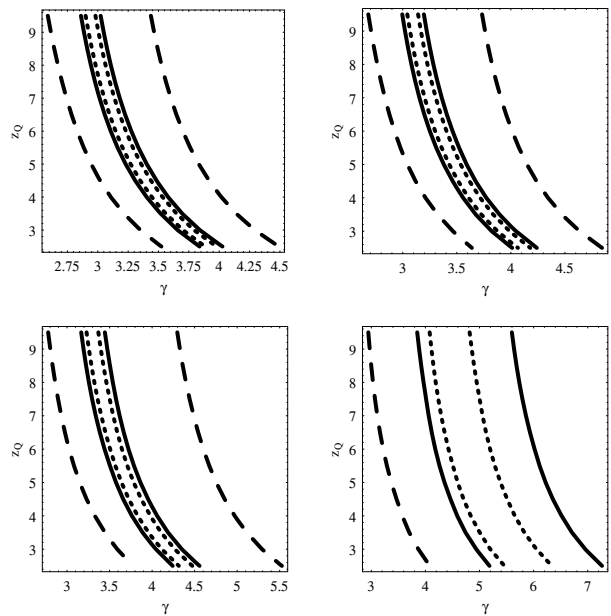}}
\caption{Level contours for $\Delta D$ in the plane $(\gamma,
z_Q)$ at $z = 0$ (top left), $0.15$ (top right), $0.35$ (bottom
left), and $1.0$ (bottom right). Contours are plotted for $\Delta
D = ({\pm} 0.5\%, {\pm} 1\%, {\pm} 5\%)$ with short\,-\,dashed,
solid and long\,-\,dashed lines, respectively. Note that, for $z =
1.0$, the deviations are so small that the contour line at $\Delta
D = 5\%$ lies outside the plot.} \label{fig: figtwo}
\end{figure}

Figure \ref{fig: figone} also shows that, for a fixed $a$, the
behavior of $\Delta D$ with $z_Q$ depends on what the value of
$\gamma$ is. For instance, a better agreement with the
$\Lambda$CDM model prediction is achieved for higher $z_Q$ if
$\gamma \le 4.5$, while the opposite is true for $\gamma \ge 7.5$.
Actually, $\Delta D$ turns out to depend only weakly on $z_Q$ for
$\gamma > 4.5$ so that it is this latter parameter that mainly
determines the behavior of the growth factor $D$ with $a$ in such
a regime. To better investigate how $\Delta D$ depends on
$(\gamma, z_Q)$, it is therefore interesting to look at the
contours of equal $\Delta D$ in the $(\gamma, z_Q)$ plane, which
are plotted in Fig.\,\ref{fig: figtwo} for some representative
values of the redshift $z$. Consider, for instance, the results
for $z = 0.15$ (top right panel in Fig.\,\ref{fig: figtwo}). To
have $|\Delta D| \le 5\%$, larger values of $z_Q$ are markedly
preferred only if coupled with low values of $\gamma$, while $z_Q
\sim 3$ is allowed, provided that $\gamma$ stays in the range
$(3.7, 4.7)$. As a general rule, the lower the value of $z_Q$ is,
the $\gamma$ higher must be to still have $|\Delta D| \le 5\%$.
With this caveat in mind, we note, however, that, unless one
chooses $z_Q > 5$ (which is rejected by the SNeIa fit), $\gamma
\le 3$ is disfavored by the requirement that the evolution of
density perturbations in the inflessence model mimicks that in the
$\Lambda$CDM one within $5\%$ over the range $0 \le z \le 1$. In
particular, remembering Eq.(\ref{eq: wlimbis}), we argue that
models in which the universe ends with a Big Rip are preferred.

It is interesting to note that fitting the SNeIa Hubble diagram
with priors on the shift and acoustic peak parameters gives $3.17
\le \gamma \le 5.86$ and $z_Q \le 5.3$ at the $95\%$ confidence
level (\cite{ie}). Although the best fit values $(\gamma, z_Q) =
(3.73, 0.1)$ are likely to be excluded because of the large values
of $\Delta D$, it is nevertheless possible to find values of
$(\gamma, z_Q)$ that make it possible both to fit the kinematic
data and give rise to an evolution of the structure as similar as
possible to that in the concordance $\Lambda$CDM model.

It is worth noting that the above results could be qualitatively
explained considering the properties of the inflessence fluid. As
explained in the previous section, the energy density tracks that
of matter (i.e., $\rho \propto a^{-3}$) for $z_Q << z << z_I$,
while, for $z << z_Q$, $\rho \propto (1 + z)^{3 - \gamma}$. A
large value of $z_Q$ means that the tracking of matter is achieved
only at high redshift when the dark energy density has become
negligible with respect to that of matter. For this same reason,
high values of $\gamma$ are preferred since it allows $\rho_X$ to
increase with $z$ at a slower rate with respect to $\rho_M$. As a
general rule, the preferred values of $(\gamma, z_Q)$ are those
that render the inflessence energy density negligible with respect
to the matter density during the structure formation epoch. This
is the same mechanism achieved in the concordance $\Lambda$CDM
model, thus explaining the shape of the contours of equal $\Delta
D$ in the $(\gamma, z_Q)$ plane.

\begin{figure}
\centering \resizebox{8.5cm}{!}{\includegraphics{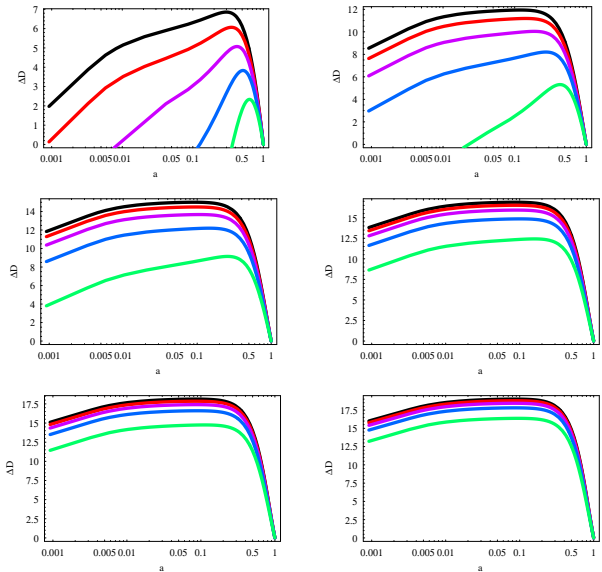}}
\caption{Same as Fig.\,\ref{fig: figone}, but now $D$ is
normalized to $D_{\Lambda}$ at the present day and the curves in
each panel refer to $z_Q$ from 1.0 to 5.0, from the lowermost to
the uppermost.} \label{fig: figonenorm}
\end{figure}

The growth factor $D$ could also be normalized to the present day
value for the $\Lambda$CDM model, i.e., by setting $D(1) =
D_{\Lambda}(1)$. Such an approach may be motivated considering
that structure formation is in remarkably good agreement with the
observations on the low redshift large\,-\,scale structure of the
universe. By normalizing to this model at present, we may better
investigate how the growth factor deviates from the $\Lambda$CDM
one in the past. Figure \ref{fig: figonenorm} shows $\Delta D$ (as
defined above) as a function of $a$ for different values of the
model parameters $(\gamma, z_Q)$. Comparing Figs.\,\ref{fig:
figone} and \ref{fig: figonenorm}, we immediately see that now
$\Delta D$ is positive (i.e., $D(a) < D_{\Lambda}$) over the full
range explored, whatever the values adopted for $(\gamma, z_Q)$
are. Moreover, $\Delta D$ is an increasing function of both
$\gamma$ and $z_Q$ for a given $a$, although the dependence on
$z_Q$ turns out to be quite weak for larger values of $\gamma$. To
better investigate how $\Delta D$ depends on the model parameters,
we plot the contours of equal $\Delta D$ in Fig.\,\ref{fig:
figtwonorm} for some representative values of the redshift $z$. As
expected, at low redshift, the deviations are so small that a
large part of the parameter space gives rise to a growth factor
consistent (to well within $\sim 5\%$) with the one for the
$\Lambda$CDM model, so that it is likely that structure formation
takes place in the same way. Not surprisingly, such a region
shrinks as the redshift increases as $\Delta D$ gets higher.
Nevertheless, at $z = 1000$, $\Delta D < 15\%$ over almost the
full parameter space considered.

\begin{figure}
\centering \resizebox{8.5cm}{!}{\includegraphics{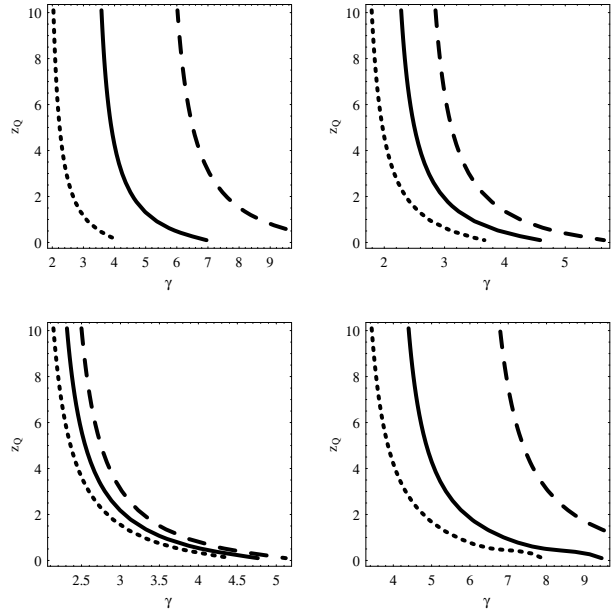}}
\caption{Level contours for $\Delta D$ in the plane $(\gamma,
z_Q)$ at $z = 0.15$ (top left), $0.35$ (top right), $1.0$ (bottom
left), and $1000$ (bottom right). Contours are plotted for $\Delta
D = (1\%, 2\%, 3\%)$ with short\,-\,dashed, solid, and
long\,-\,dashed lines, respectively, in the first three panels,
while $\Delta D = (5\%, 10\%, 15\%)$ are used for $z = 1000$.}
\label{fig: figtwonorm}
\end{figure}

\begin{table}
\begin{center}
\begin{tabular}{|c|c|c|c|c|}
\hline
$\Omega_M, \gamma, z_Q$ & $\alpha$ & $d_1$ & $d_2$ & $d_3$ \\
\hline  \hline $0.28, 3.7, 0.1$ & $0.1355$ & $-0.5725$ & $-0.0471$
& $-0.0124$ \\ $0.24, 5.0, 4.0$ & $0.3371$ & $-0.6510$ & $-0.0878$
& $-0.0367$ \\ $0.24, 5.5, 5.0$ & $0.3474$ & $-0.6275$ & $-0.0814$
& $-0.0377$ \\ $0.24, 6.0, 7.0$ &
$0.3513$ & $-0.5983$ & $-0.0731$ & $-0.0372$ \\
\hline
\end{tabular}
\end{center}
\caption{Values of the fitting parameters in Eq.(\ref{eq:
approxd}) for some representative sets of $(\Omega_M, \gamma,
z_Q)$. The first row refers to the best fit to SNeIa fit, while
the remaining rows refer to models providing a good fit to the
CMBR anisotropy spectrum.}
\end{table}

Although an analytical solution of Eq. (\ref{eq: eqpertd}) is not
available and numerical integration is straightforward, we believe
it is useful to have an approximate expression for $D(a)$ to be
used in data fitting. To this aim, we have integrated Eq.
(\ref{eq: eqpertd}) for $a_{LS} \le a \le 1$, and found that a
very good approximation is given by the fitting formula\,:

\begin{equation}
D(a) \simeq d_0 {\rm e}^{\alpha \eta} \left (1 + d_1 \eta + d_2
\eta^2 + d_3 \eta^3 \right ) \ , \label{eq: approxd}
\end{equation}
with $\eta = \ln{a}$, $d_0 = D(a = 1)$, and $(\alpha, d_1, d_2,
d_3)$ constant parameters depending on $(\Omega_M, \gamma, z_Q)$.
Equation (\ref{eq: approxd}) works very well with an rms error
smaller than $2\%$ up to $z \simeq 400$ and less than $8\%$ at $z
= 1100$. Unfortunately, we have not been able to find satisfactory
approximated formulae for $(\alpha, d_1, d_2, d_3)$ in terms of
$(\Omega_M, \gamma, z_Q)$, although we have generated tables
(available on request to the authors) for $0.05 \le \Omega_M \le
0.55$, $2.5 \le \gamma \le 7.5$ and $0.1 \le z_Q \le 7.1$, which
may be easily interpolated to get the corresponding fitting
parameters. As an example, we report their values in Table\,1 for
some interesting cases.

\section{Growth index}

A quantity that can be measured by the galaxy correlation function
or the peculiar velocities is the so\,-\,called {\it growth index}
defined as\,:

\begin{equation}
f \equiv \frac{d\ln{\delta}}{d\ln{a}} = \frac{a}{\delta}
\frac{d\delta}{da} \ . \label{eq: defgi}
\end{equation}
From a theoretical point of view, to estimate $f$ for a given dark
energy model, one may solve Eq. (\ref{eq: pertscale}) to get
$\delta = \delta(a)$ and then straightforwardly compute $f$.
However, since we often have to deal with numerical integration,
it is better to directly solve an equation for $f$, to avoid
propagating the numerical errors. To this end, one should simply
use definition (\ref{eq: defgi}) of $f$ and multiply Eq. (\ref{eq:
pertscale}) by $a/\delta$ to finally get the evolution equation
for the growth index\,:

\begin{equation}
f^{\prime} + \frac{f^2}{a} + \left [ \frac{2}{a} +
\frac{(\ln{E^2})^{\prime}}{2} \right ] f - \frac{3 \Omega_M}{2 E^2
a^4} = 0 \ . \label{eq: eqf}
\end{equation}
Not surprisingly, Eq. (\ref{eq: eqf}) must be integrated
numerically, using $f(a_{LS}) = 1$ as the initial condition, and
Eqs. (\ref{eq: ea}) and (\ref{eq: ga}) for the particular case of
the inflessence model.

\begin{figure}
\centering \resizebox{8.5cm}{!}{\includegraphics{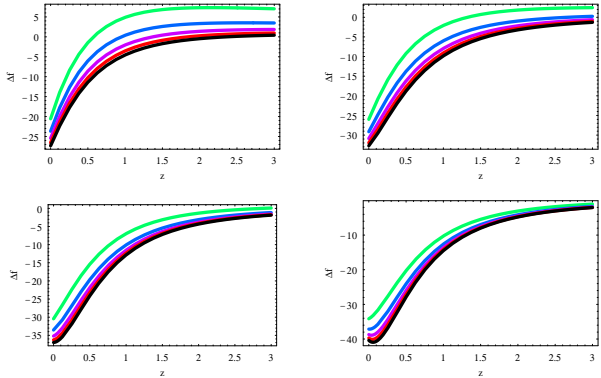}}
\caption{$\Delta f$ vs. $z$ for models with $\gamma = 3.5$ (top
left), $4.5$ (top right), $5.5$ (bottom right), and $6.5$ (bottom
right) and $z_Q$ from $1.0$ to $5.0$ (in steps of $1$) from the
uppermost to the lowermost curves. The other parameters are set as
explained in the text.} \label{fig: figtrheenew}
\end{figure}

The 2dFGRS collaboration measured the position and the redshift of
over 220000 galaxies and, from the analysis of the correlation
function, determined the redshift distortion parameter $f/b$ with
the bias parameter $b$ quantifying the difference between the
galaxies and dark haloes distributions. Using the estimated $f/b$
and the two different methods employed by Verde et al. (2001) and
Lahav et al. (2002) to determine the bias $b$, one may estimate $f
= 0.51 {\pm} 0.1$ or $f = 0.58 {\pm} 0.11$ at the survey effective
depth $z = 0.15$.

Both of these estimates are in very good agreement with what is
predicted by the $\Lambda$CDM model, so that it is interesting to
compare the behaviour of $f$ predicted by the inflessence model
with that of the concordance scenario. To this end, we define
$\Delta f = 1 - f/f_{\Lambda}$, which gives the percentage
difference between the predictions of the two models. This is
shown in Fig.\,\ref{fig: figtrheenew}, multiplied by 100 for the
sake of clarity, considering different values of $\gamma$ and
$z_Q$ and setting the other parameters as in Sect.\,3. It is worth
noting that $\Delta f$ is always negative, i.e. the growth index
$f$ of the inflessence model is larger than the $\Lambda$CDM one
over the whole parameter space $(\gamma, z_Q)$. It is therefore
mandatory to directly compare $f(z = 0.15)$ with the observed
value (which is in good agreement with the concordance model
predictions) to check whether the overestimate of $f$ may be
troublesome thus allowing us to put constraints on the parameter
space.

\begin{figure}
\centering \resizebox{8.5cm}{!}{\includegraphics{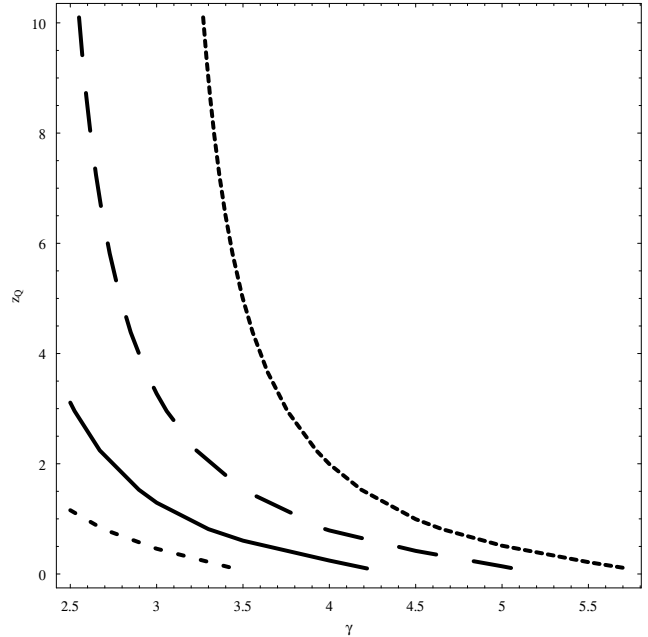}}
\caption{Level contours for $\Delta f$ at $z = 0.15$ in the plane
$(\gamma, z_Q)$. Contours are plotted for $\Delta f(z = 0.15) =
-5\%$ (short\,-\,dashed), $-10\%$ (solid), and $-15\%$
(long\,-\,dashed). Models lying to the left of the dotted line
have $w_0 \ge -1$, while the ones to the right behave today as
phantoms, i.e., $w_0 < -1$.} \label{fig: figfournew}
\end{figure}

As in the case of $\Delta D$, the dependence of $\Delta f$ on the
model parameters $(\gamma, z_Q)$ is particularly involved, so that
it is better to look at the contours of equal $\Delta f$ in the
plane $(\gamma, z_Q)$. These are shown in Fig.\,\ref{fig:
figfournew}, where we have set $z = 0.15$ to compare with the
observed value of $f$ measured by the SDSS survey. It is worth
noting that such a plot could be used to constrain $(\gamma, z_Q)$
by requiring that $\Delta f(z = 0.15)$ be lower than a given
threshold dictated by the estimated $f$. It is clear that such a
method has the potential to severely narrow the region of the
parameter space $(\gamma, z_Q)$ in agreement with the
observations. Moreover, a comparison of Fig.\,\ref{fig:
figfournew} with the projected likelihood contours from the SNeIa
Hubble diagram fitting shows that they are orthogonal, so that a
combined analysis may place strong constraints on the inflessence
model parameters. Unfortunately, this method is still not
applicable at the moment because the estimated $f$ is still
affected by a large percentage error ($\sim 20\%$), which makes
the test useless, since, as Fig.\,\ref{fig: figfournew} shows,
$\Delta f \le 15\%$ over a wide region of the $(\gamma, z_Q)$
plane. This is essentially due to the low redshift tested ($z =
0.15$), but extending the measurement to higher $z$ (also with the
same percentage error) could significantly improve the efficiency
of such an analysis. Nevertheless, it is worth noting that the
allowed region of the parameter space $(\gamma, z_Q)$ lies to the
left of the $w = -1$ line so that phantom like models are
excluded. Such a result is nicely consistent with the constraints
from the SNeIa fit that also point towards this conclusion.

Using the definitions of $D$ and $f$, one easily gets\,:

\begin{equation}
f = D \left ( 1 + \frac{1}{D} \frac{d\ln{D}}{d\ln{a}} \right )
\label{eq: fvsd}
\end{equation}
so that $f$ not only depends on how density perturbations evolve,
but also on their logarithmic rate of evolution. As such, $f$ is a
more subtle quantity depending on both the expansion history and
the structure evolution. As a general remark, we note that $\Delta
f$ approaches null values more quickly than $\Delta D$ as a
consequence of its logarithmic nature. It is nevertheless
interesting to compare the contours of equal $\Delta D$ with those
of equal $\Delta f$ to see whether narrower constraints on the
model parameters $(\gamma, z_Q)$ may be obtained by imposing the
same threshold on both quantities. Indeed, although the contours
turn out to be parallel, they are also shifted towards each other
so that the overlapping region is significantly narrower than
those selected by the constraint on $\Delta f$ alone. Should we
have an observationally motivated constraint on $\Delta D$ (as
that on $\Delta f$) to the $5\%$ level of precision, we could thus
efficiently constrain the inflessence parameters $(\gamma, z_Q)$.

Indeed, it is likely that future measurements (using a larger
redshift survey observing more galaxies) should lessen the error
on the observed $f(z = 0.15)$ to the $5\%$ level. In such a case,
it would be useful to have an approximated formula for $f(z =
0.15)$ as a function of the inflessence model parameters. This is
given as\,:

\begin{equation}
f(z = 0.15) \simeq f_0 \Omega_M^{f_1 + f_2 \ln{\gamma} + f_3
\ln{z_Q}} \label{eq: approxf}
\end{equation}
with\,:

\begin{displaymath}
(f_0, f_1, f_2, f_3) = (0.8778, 0.4614, -0.1839, -0.0323) \ ,
\end{displaymath}
which works quite well (with an rms error less than $1.5\%$) for
$0.2 \le \Omega_M \le 0.4$, $3 \le \gamma \le 6$ and $1 \le z_Q
\le 7$. It is worth noting that, for the $\Lambda$CDM model, one
has the approximated formula $f_{\Lambda}(z) \simeq \left [
\Omega_M (1 + z)^{3} \right ]^{0.55}$ (\cite{SW94}; \cite{WS98};
\cite{LBH04}). For the inflessence model, we find a similar
formula, but the value of the exponent is determined by the model
parameters $(\gamma, z_Q)$. Note, however, that Eq. (\ref{eq:
approxf}) only holds for $z = 0.15$. Although we have not
explicitly checked it, it is likely that the same formal
expression holds over a large range in $z$, provided that the
parameters $(f_0, f_1, f_2, f_3)$ are suitably changed.

\begin{figure}
\centering \resizebox{8.5cm}{!}{\includegraphics{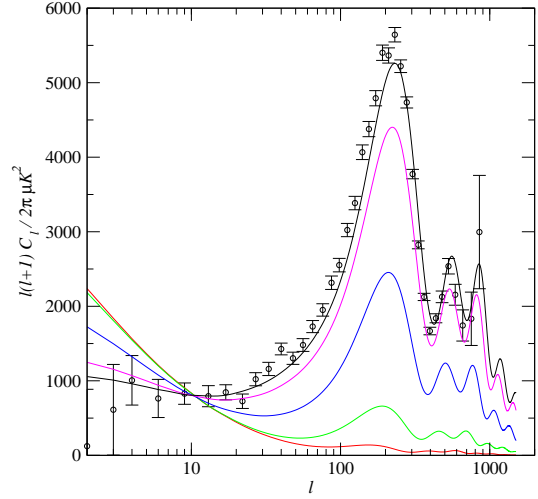}}
\caption{The CMBR anisotropy spectrum for the inflessence model
with $z_Q = 0.1$ and $\gamma = 3, 4, 5, 6, 7$ (from bottom to
top). Other parameters are set as explained in the text. Data
points are the WMAP measurements.} \label{fig: zqcl}
\end{figure}

\section{CMBR anisotropy spectrum}

Since its discovery by Penzias and Wilson (1965), the CMBR has
played a fundamental role in cosmology. The recent precise
measurement of its anisotropy spectrum by the WMAP collaboration
(\cite{WMAP}) has further increased the importance of such an
observable in assessing the viability of any cosmological model.
Unfortunately, the large number of parameters entering the
determination of the anisotropy spectrum makes it quite difficult
to extract constraints on a given model's parameters from a
time\,-\,expensive likelihood analysis (typically based on a Monte
Carlo Markov Chain exploration of the wide parameter space). This
is also the case for the inflessence model, so that we will only
investigate how the spectrum changes as a function of the main
model parameters, i.e., $(\gamma, z_Q)$. In the following
analysis, we therefore set the following values for the other
parameters involved in the computation\,:

\begin{figure}
\centering \resizebox{8.5cm}{!}{\includegraphics{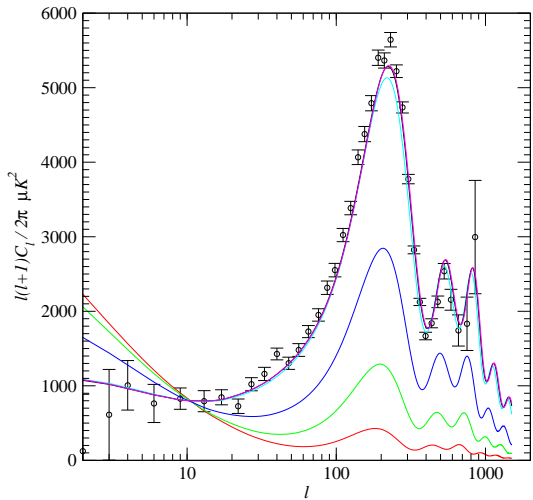}}
\caption{The CMBR anisotropy spectrum for the inflessence model
with $\gamma = 3.73$ and $z_Q = 0.1, 0.5, 1, 3, 5$ (from bottom to
top). Other parameters are set as explained in the text. Data
points are the WMAP measurements.} \label{fig: gammacl}
\end{figure}

\begin{figure}
\centering \resizebox{8.5cm}{!}{\includegraphics{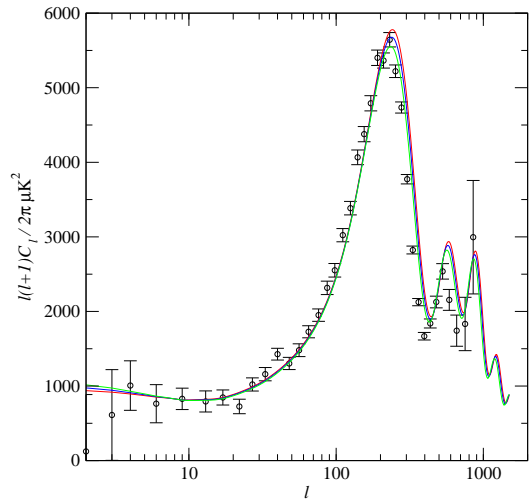}}
\caption{The CMBR anisotropy spectrum for the inflessence model
with $(\gamma, z_Q) = (6, 7)$, $(5.5, 5)$, and $(5, 4)$, from top
to bottom. Other parameters are set as explained in the text. Data
points are the WMAP measurements.} \label{fig: clok}
\end{figure}

\begin{displaymath}
\omega_b = 0.04 \ , \ \Omega_M = 0.24 \ , \ h = 0.70 \ , \ \tau =
0.17 \ , \ n = 1
\end{displaymath}
with the optical depth $\tau$ and the spectral index (assuming no
running) $n$. The remaining inflessence parameters $(\beta, z_I)$
are set to their fiducial values $(-3, 3454)$. To compute the CMBR
anisotropy spectrum, we use CMBFAST (\cite{SZ96}), obtaining the
results plotted in Figs.\,\ref{fig: zqcl} and \ref{fig: gammacl},
where the inflessence parameters $(\gamma, z_Q)$ has been chosen
with the main aim of highlighting the dependence on these
parameters rather than providing a good fit to the data.

As in the case of the growth factor $D$, large values of both
$\gamma$ and $z_Q$ are needed to get satisfactory results. Indeed,
while the position of the peaks is essentially independent on the
parameters $(\gamma, z_Q)$, their amplitude is an incresing
function of these two quantities. Moreover, at the low multipoles
($l \le 10$), the spectrum is significantly overestimated for
small values of the inflessence parameters. It is worth stressing,
however, that very good fits may be achieved by suitably tuning
the parameters $(\gamma, z_Q)$. Some nice examples are shown in
Fig.\,\ref{fig: clok}.

It is worth noting that such large values of $(\gamma, z_Q)$ stay
at the upper end of the confidence ranges obtained from the
likelihood analysis performed in Cardone et al. (2005).
Nevertheless, they are not excluded, so that we are confident that
a combined analysis of the whole parameter space (thus also
changing the matter density parameter $\Omega_M$ that we have held
fixed up to now, the spectral index $n$, and the optical depth
$\tau$) could pinpoint a narrow region in the parameter space
giving a satisfactory fit to SNeIa Hubble diagram, growth index,
and CMBR anisotropy spectrum.

\section{Conclusions}

The inflessence model has been proposed as a possible mechanism to
explain both the inflationary epoch in the early universe and the
present day cosmic speed up. According to this scenario, a single
fluid with the energy density given by Eq. (\ref{eq: rhovsa}) is
added to radiation and dust matter, thus working as the inflaton
field at very low $a$ (i.e., for $z >> z_I$) and as dark energy on
the scale $a \sim 1$ (namely, for $z \le z_Q$). Since this model
has been shown to be able to nicely fit the SNeIa Hubble diagram,
while also giving correct values for the shift and acoustic peak
parameters, it is worth wondering how structure formation takes
place.

To this end, we have investigated the evolution of density
perturbations in the linear regime, comparing both the growth
factor $D(a)$ and the growth index $f(z)$ to these same quantities
in the $\Lambda$CDM model. In particular, we have concentrated our
attention on the two inflessence parameters $\gamma$ and $z_Q$,
which determine, respectively, the asymptotic value of the eos
(and hence the final fate of the universe) and the transition from
the matter like to the quintessence like scaling of the energy
density with $a$. Moreover, since $\gamma$ and $z_Q$ also set the
present day value of the EoS, one can easily understand that they
play a leading role in determining both $D(a)$ and $f(z)$. As a
further test, we have also computed the CMBR anisotropy spectrum
for fixed values of the other parameters (especially the optical
depth $\tau$ and the spectral index $n$).

As a general result, we have found that, using large values of
$\gamma$ and $z_Q$, it is possible to work out scenarios in which
structure formation takes place in quite similar ways in both the
inflessence and the $\Lambda$CDM models. Moreover, for these same
values, the predicted CMBR anisotropy spectrum also nicely agrees
with the WMAP data. Such large values seem to be disfavored by the
fitting to the SNeIa Hubble diagram, so that some tension between
these two different probes is present. However, it is worth noting
that the constraints coming from SNeIa are rather weak so that it
is indeed possible that such a conflict is not particularly
worrisome.

It is also worth noting that a precise determination of the growth
index $f$ (at the $5 - 10\%$ level) at the  low redshift typical
of present day galaxy surveys or a measurement of $f$ at a higher
redshift have the potential to severely constrain the parameters
$(\gamma, z_Q)$. Moreover, such constraints are orthogonal to
those coming from SNeIa, so that a joint analysis could
definitively assess the viability of the inflessence model and
pinpoint a narrow range in the parameter space $(\Omega_M, \gamma,
z_Q)$. One could also include the CMBR anisotropy spectrum in a
fully comprehensive likelihood test. However, such an approach is
likely to be affected by strong degeneracies among the five
inflessence parameters $(\Omega_M, \beta, \gamma, z_I, z_Q)$ and
the other CMBR parameters such as the optical depth $\tau$, the
baryon content $\omega_b$, and the spectral index $n$ (and its
eventual running $dn/d\ln{k}$). To probe such a large parameter
space, a Monte Carlo Markov Chain approach is mandatory and is
left for future works.

Actually, having determined the growth factor $D(a)$, for which we
have also found an analytical approximation, we may further
explore the issue of structure formation in the inflessence model.
To this aim, one could use the estimated $D(a)$ to estimate the
critical overdensity for collapse at the present day, and as a
function of time, hence determining the mass function through the
Press \& Schechter formalism (\cite{PS74}) for the spherical
collapse of perturbations, or its generalization to elliptical
collapse worked out by Sheth and Tormen (1999). The mass function
is the key ingredient to predicting cluster number counts, which
are known to be a powerful test of dark energy models (see, e.g.,
\cite{HMP01}). In many observational applications, it is also
interesting to check whether the collapsed perturbation is
virialized or not. As has been pointed out in Percival (2005),
dark energy also plays a role in this process, and hence it is
interesting to investigate how the inflessence model affects this
important process. Most of these problems will be addressed in a
forthcoming paper.

\end{document}